\documentclass[iop]{emulateapj}

\begin{document}
\title{Turbulence in the TW Hya disk}
\author{Kevin M. Flaherty\altaffilmark{1}
A. Meredith Hughes\altaffilmark{1}
Richard Teague\altaffilmark{2}
Jacob B. Simon\altaffilmark{3,4}
Sean M. Andrews\altaffilmark{5}
David J. Wilner\altaffilmark{5}
}
\altaffiltext{1}{Van Vleck Observatory, Astronomy Department, Wesleyan University, 96 Foss Hill Drive, Middletown, CT 06459}
\altaffiltext{2}{Department of Astronomy, University of Michigan, 500 Church St., Ann Arbor, MI 48109}
\altaffiltext{3}{Department of Space Studies, Southwest Research Institute, Boulder, CO 80302}
\altaffiltext{4}{JILA, University of Colorado and NIST, 440 UCB, Boulder, CO 80309}
\altaffiltext{5}{Harvard-Smithsonian Center for Astrophysics, 60 Garden Street, Cambridge, MA 02138}

\begin{abstract}
Turbulence is a fundamental parameter in models of grain growth during the early stages of planet formation. As such, observational constraints on its magnitude are crucial. Here we self-consistently analyze ALMA CO(2-1), SMA CO(3-2), and SMA CO(6-5) observations of the disk around TW Hya and find an upper limit on the turbulent broadening of $<$0.08c$_s$ ($\alpha<$0.007 for $\alpha$ defined only within 2-3 pressure scale heights above the midplane), lower than the tentative detection previously found from an analysis of the CO(2-1) data. We examine in detail the challenges of image plane fitting vs directly fitting the visibilities, while also considering the role of the vertical temperature gradient, systematic uncertainty in the amplitude calibration, and assumptions about the CO abundance, as potential sources of the discrepancy in the turbulence measurements. These tests result in variations of the turbulence limit between $<$0.04c$_s$ and $<$0.13c$_s$, consistently lower than the 0.2-0.4c$_s$ found previously. Having ruled out numerous factors, we restrict the source of the discrepancy to our assumed coupling between temperature and density through hydrostatic equilibrium in the presence of a vertical temperature gradient and/or the confinement of CO to a thin molecular layer above the midplane, although further work is needed to quantify the influence of these prescriptions. Assumptions about hydrostatic equilibrium and the CO distribution are physically motivated, and may have a small influence on measuring the kinematics of the gas, but they become important when constraining small effects such as the strength of the turbulence within a protoplanetary disk.



\end{abstract}

\section{Introduction}
Turbulence is an important parameter in the planet formation process. For micron to cm sized dust grains, turbulence influences their vertical settling \citep[e.g.][]{dub95,joh05,cie07,you07}, radial diffusion \citep{cla88,des17}, collisional growth \citep{orm07,bir10}, and trapping in asymmetric vortices \citep{pin15}. Turbulent eddies may encourage the rapid creation of planetesimals through the concentration of the micron to cm sized grains \citep{bar95,kla97,joh07,cuz08}, while the efficient dust settling associated with weak turbulence is a key factor in the creation of planetesimals through the streaming instability \citep[e.g.][]{you05} and planetesimal growth due to pebble accretion \citep{xuz17}. During the later stages of planet formation, turbulence affects the collisional velocities of planetesimals \citep{ida08}, the radial migration of sub-Jupiter mass planets \citep{nel04,ois07,bar11} and the ability of more massive planets to open a gap \citep[e.g.][]{kle12,fun14}. Chemical mixing of the gas will depend on the turbulence \citep{sem11}, which in turn can influence the abundance of common gas species such as CO \citep{fur14,xur17}. This is not to mention the global evolution in the gas surface density over the lifetime of of the disk that is a result of angular momentum transport associated with turbulence \citep[e.g.][]{lyn74}, or the turbulence-dependent ability of certain processes to create large scale structures in the disk \citep[e.g.][]{bae17}.

A number of turbulence generating mechanisms have been studied, including the Vertical Shear Instability \citep{nel13,lin15}, the Zombie Vortex Instability \citep{mar15,les16}, Gravitational Instabilities \citep{gam01,for12,hir17}, and Baroclinic Instabilities \citep{kla03,lyr11}, with a leading candidate for generating turbulence in protoplanetary disks being the Magneto-Rotational Instability \citep{bal98,fro06,sim13,sim15,bai15}. While a great deal of theoretical effort has been expended in determining how turbulence operates in the cold, dense disks surrounding young stars, few observational constraints exist. Much of this difficulty lies in the fact that turbulence is buried beneath the stronger Keplerian and thermal motions of the disk. Keplerian velocities around a solar-type star range from 100 km s$^{-1}$ to 2 km s$^{-1}$ at distances of 0.1-200 au, while the thermal motion of H$_2$ can vary from 3 km s$^{-1}$ - 0.3 km s$^{-1}$ from T=2000 K in the inner disk to 20 K in the outer disk. Turbulence produces motions that are, at the most, equivalent to the thermal motion, requiring precise measurements of the disk motion to below 0.1 km s$^{-1}$. Such modest velocities are comparable in size to secondary effects on the dominant kinematics, such as the variation in Keplerian velocity with height above the midplane and the contribution of the pressure gradient \citep{ros13}, complicating attempts to measure turbulence.


Observations of spatially resolved molecular line emission shows promise for revealing turbulent motions in the outer disk. Early work using the Submillimeter Array and the Plateau de Bure Interferometer focused on high spectral resolution observations, designed to measure the spectral broadening associated with non-thermal motion. The use of relatively heavy molecules helps reduce the thermal motion, making it easier to find the non-thermal motion; for 20 K gas the isothermal sound speed (=$\sqrt{2k_BT/\mu m_h}$) is 0.26 km s$^{-1}$, but the thermal broadening of CO (=$\sqrt{2k_BT/m_{CO}}$) is only 0.08 km s$^{-1}$, which is equivalent to 30\%\ of the local sound speed. \citet{hug11} examined 44 m s$^{-1}$ resolution observations of CO(3-2) from the disks around HD 163296 and TW Hya, finding a marginal detection in HD 163296 and an upper limit in TW Hya. \citet{gui12} report detectable turbulence around DM Tau of 0.4-0.5c$_s$. Historically sub-mm observations have been most sensitive to the outer ($>$50 au) disk, while the inner disk is more directly probed with infrared spectroscopy. \citet{car04}, in examining R$\sim$25,000 near infrared observations of water in the inner ($<$0.3 au) disk around SVS 13, require an additional source of broadening beyond the thermal and Keplerian motion of the disk, consistent with trans-sonic turbulence. These results, taken at face value, suggest a wide range of turbulent velocities between different protoplanetary disk systems and different radial and vertical locations within the disk. 

The greater sensitivity and angular resolution of ALMA have enabled far more precise measurements of the line widths of molecular emission at millimeter wavelengths. \citet{fla15,fla17} examine the disk around HD 163296 using a mix of molecular tracers (CO(3-2), CO(2-1), $^{13}$CO(2-1), C$^{18}$O(2-1), and DCO$^+$(3-2)) and find upper limits on the turbulent broadening of $<$5\% of the local sound speed. \citet{tea16} (hereafter T16) analyze the disk around TW Hya using observations of CO(2-1), CS(5-4), and CN(2-1) and report turbulence of 0.2-0.4 c$_s$ for CO(2-1), with marginal detections of turbulent broadening from CS and CN. According to these measurements, the disk around TW Hya appears to be substantially more turbulent than the disk around HD 163296. 

While there may be physical factors that can explain the different levels of turbulence between HD 163296 and TW Hya (e.g. strength of the vertical magnetic field, FUV ionization), we must first examine the role of differences in methodologies on the results. \citet{fla15,fla17} employ a parametric disk model that self-consistently calculates the temperature and density structure, which is fed through a ray-tracing radiative transfer code to generate model visibilities that are compared directly to the data. T16 fit high resolution spectra either by considering each pixel in the image independently or by assuming that temperature and turbulence follow radial power law structures. Here we analyze the ALMA observations of TW Hya that were first presented in T16, along with archival observations of CO(3-2) and CO(6-5), with the methodology used to examine HD 163296 in \citet{fla17}. By studying a similar data set, we can isolate differences in methodology. In section 3 we present the model structure that has previously been used to constrain turbulence in the disk around HD 163296. In Section 4 we discuss our finding of an upper limit on the turbulence that falls below the measurement derived by T16, and explore possible sources of the discrepancy between our results and those of T16. We consider in detail the effects of systematic uncertainty, image plane vs visibility domain fitting, as well as assumptions about the thermal structure and CO abundance. We consider the implications of the modest turbulence in Section 5.

\section{Observations\label{data}}
To directly compare our method for deriving turbulent line broadening in the TW Hya disk with the results of T16, we extract the CO(2-1) data from project 2013.1.00387.S (PI: S. Guilloteau) that was used in their analysis. The visibilities were calibrated using the standard ALMA calibration script and were self-calibrated using the continuum spectral window. The spectral window containing the CO(2-1) line has 15 kHz (40 m s$^{-1}$) channels and baselines extending from 16-424 k$\lambda$ (20-550 meters). Imaging the line emission with the MIRIAD clean task, using robust=0.5 visibility weighting results in a beam size of 0$\farcs$64x0$\farcs$51, which is larger than the 0$\farcs$50x0$\farcs$42 beam size of T16, generated with the CASA clean task with a robust of 0.5. This discrepancy may be partly due to errors in antenna positions that had been present in the data set originally provided to the PI, but were corrected by the time the data became publicly available. The phase errors introduced by these incorrect antenna positions were corrected by T16 through self-calibration, leading to improved S/N of their data. Subsequent differences in the relative weighting of the baselines may also contribute to the different beam sizes. Accounting for this difference in beam size, the flux in our images is within the uncertainties of those generated by T16.

We also analyze archival SMA observations of CO(3-2) \citep{hug11} and CO(6-5) \citep{qi06}. Since these emission lines arise from different upper energy states, they are valuable for constraining the gas temperature, as discussed below. The CO(3-2) observations were taken in 2008 with baselines of 16-182m, corresponding to a spatial resolution of 1$\farcs$0x0$\farcs$8, and a spectral resolution of 44 m s$^{-1}$. The CO(6-5) observations were conducted in 2005 with baselines ranging from 7 to 68m, corresponding to a synthesized beam of 3$\farcs$9 x 1$\farcs$2, and a velocity resolution 350 m s$^{-1}$. Details on the reduction and sensitivity of these data are available in \citet{hug11} and \citet{qi06} respectively. While we analyze all three data sets simultaneously, the large visibility weights of the ALMA observations, due to the low noise in these data, will place a strong preference on the regions of parameter space that best accommodates the ALMA data.

\section{Parametric models\label{model}}
The modeling code used to match the observations (CO(2-1)/CO(3-2)/CO(6-5) at $\nu$=230.538 GHz, 345.79599 GHz, and 691.47308 GHz respectively) was used previously in \citet{fla15} and \citet{fla17} and is based on earlier work by \citet{ros13} and \citet{dar03}. The basic equations, and definitions of important parameters, are listed below.

The temperature structure is defined as a power law with radius, with a vertical gradient connecting the cold midplane with the warm atmosphere. 
\begin{eqnarray}
T_{\rm mid} = T_{\rm mid0}\left(\frac{r}{150\ \rm au}\right)^{q_{mid}}\\
T_{\rm atm} = T_{\rm atm0}\left(\frac{r}{150\ \rm au}\right)^{q_{atm}}\\
T_{\rm gas}(r,z) = \left\{
\begin{array}{ll}
T_{\rm atm} + (T_{\rm mid}-T_{\rm atm})(\cos\frac{\pi z}{2Z_q})^{2} & \mbox{if $z < Z_q$} \\
T_{\rm atm} & \mbox{if $z \ge Z_q$}
\end{array}
\right.\\
Z_q = 70\ {\rm au} (r/150\ {\rm au})^{1.3}
\end{eqnarray}
This parametrization can be reduced to a vertically isothermal structure by using a single temperature normalization parameter ($T_0=T_{\rm atm0}=T_{\rm mid0}$). 

The surface density is assumed to follow a power law, with an exponential tail, as expected for a viscously evolving disk \citep{lyn74,har98}.
\begin{equation}
\Sigma_{\rm gas}(r) = \frac{M_{\rm gas}(2-\gamma)}{2\pi R^2_c}\left(\frac{r}{R_c}\right)^{-\gamma}\exp\left[-\left(\frac{r}{R_c}\right)^{2-\gamma}\right].
\end{equation}
The parameters $M_{\rm gas}$, $R_c$ and $\gamma$ are the gas mass (in M$_{\odot}$), critical radius (in au) and power law index respectively. The volume mass density is calculated using the surface density, along with the temperature, to perform the hydrostatic equilibrium calculation at each radius. The velocity field is Keplerian, with corrections for the height above the midplane and the pressure support of the gas, as in \citet{ros13}. Line broadening is treated as a Gaussian width for the line profile, with contributions from thermal motions and a non-thermal term that we associate with turbulence\footnote{While traditionally turbulence has been designated as e.g. $v_{\rm turb}$ we use the notation $\delta v_{\rm turb}$ to distinguish turbulence, as a measure of velocity dispersion, from vector quantities, such as the Keplerian orbital velocity} . We consider cases in which the non-thermal term is proportional to the local isothermal sound speed:
\begin{equation}
\Delta V = \sqrt{\left(2k_BT(r,z)/m_{CO}\right)(1+\delta v_{\rm turb}^2)},
\end{equation}
as well as if the motion is specified with a single global value in units of km s$^{-1}$:
\begin{equation}
\Delta V = \sqrt{\left(2k_BT(r,z)/m_{CO}\right) + \delta v_{\rm turb}^2}.
\end{equation}
The latter parameterization more closely matches the methodology of T16, while the former is physically motivated by numerical simulations that predict that the turbulence scales with the local sound speed \citep[e.g.][]{sim15}. CO is assumed to be distributed uniformly between a depth determined by the freeze-out at low temperatures ($T<19$ K) and a height determined by the photodissociation boundary in the disk upper atmosphere. Photodissociation is assumed to occur above vertical column densities, measured downward from the surface of the disk, of N$_{H_2}$=1.3$\times$10$^{21}$ cm$^{-2}$. Both freeze-out and photodissociation are treated as a drop in gas-phase CO abundance of eight orders of magnitude. 

\begin{figure*}
\center
\includegraphics[scale=.4]{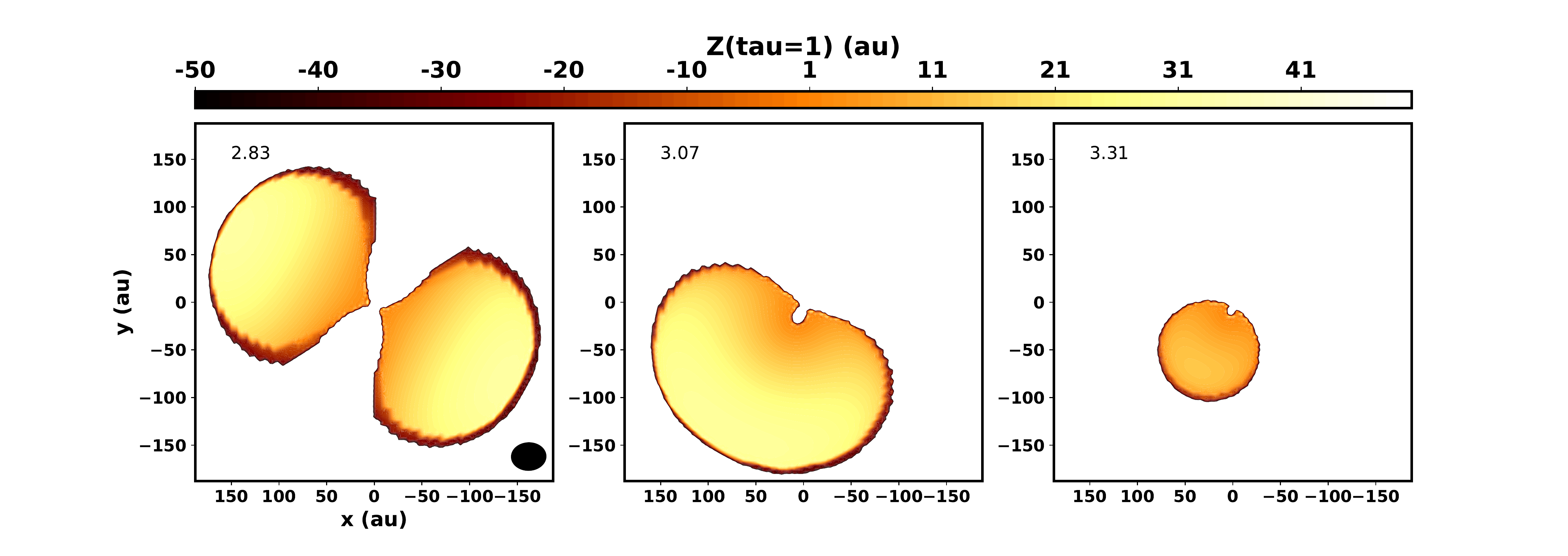}
\caption{Height above the midplane of the $\tau$=1 surface for three channels in our fiducial model; negative values corresponds to below the midplane, ie. on the far side of the disk. At the edges of the emission the optical depth is lower, often reaching the midplane, and occasionally the far side of the disk. Even for a very optically thick emission line like CO(2-1) there are optically thin lines of sight that contribute to the emission. \label{tau1_figure}}
\end{figure*}

In our fiducial model we allow $T_{\rm atm0}$, $T_{\rm mid0}$, $q_{\rm atm}$, $R_c$, inclination and $\delta v_{\rm turb}$ to vary, while fixing stellar mass, distance, CO abundance, position angle, systemic velocity, and the spatial offset of the disk center from the phase center of the observations. While CO(2-1) is optically thick, with much of the midplane obscured by emission from the upper layers, we find that the outermost edge of the emission is optically thin and pierces to the midplane (Figure~\ref{tau1_figure}). This provides a constraint on the midplane temperature at the very outer edge of the disk, but no information on the midplane temperature underneath much of the CO(2-1) emission. To guide the midplane temperature structure, we use previous measurements of the CO snow line \citep{sch16} to fix the midplane temperature at 19 K at 20 au. The parameter $q_{\rm mid}$ is allowed to vary such that $T_{\rm mid}$ passes through 19 K at 20 au and $T_{\rm mid0}$ at 150 au. 

We assume the disk mass is 0.05M$_{\odot}$, with a CO/H$_2$ abundance of 10$^{-6}$, based on the measurements of HD from \citet{ber13}. We use a stellar mass of 0.7 M$_{\odot}$, identical to that used by T16, and slightly larger than the stellar mass assumed by \citet{hug11} (0.6 M$_{\odot}$). The distance to TW Hya is assumed to be 54 pc, which is slightly smaller than the recent GAIA parallax measurement of 59.0$\pm$0.5 pc \citep{gaia16}, but is chosen to match the value used by T16. Adopting the GAIA-based distance to TW Hya would most strongly affect the temperature normalization, although the effect is small (a factor of two increase in the distance would only result in a $\sim$20\%\ larger temperature), and smaller than the range of temperatures allowed for by the amplitude calibration uncertainty, as discussed below. Using preliminary models we derive a systemic velocity of 2.825 km s$^{-1}$, the offset of the disk emission from the phase center to be [-.06\arcsec,-.04\arcsec] and a disk position angle of 151$^{\circ}$.

Model images are generated at each velocity channel using an LTE ray-tracing code and model visibilities are created using the MIRIAD task UVMODEL. The posterior distribution functions for each parameter are estimated using the MCMC routine EMCEE \citep{for13} based on the Affine-Invariant algorithm originally proposed in \citet{goo10}. The uncertainties of the visibilities at each baseline are calculated based on the dispersion in observed visibilities within line-free channels and among baselines of similar distances. Typical MCMC chains consist of 80 walkers and 1600 steps. The first 600 steps are removed as burn-in, after which the median of the posterior distributions varies by $<$1\%. We use linear spacing in all parameters except $R_c$, where we fit $\log(R_c)$. 

T16, in their pixel-by-pixel approach, exclude the inner 40 au of the disk to avoid the substantial Keplerian shear within this region. Our models fully account for the Keplerian rotation, but may instead be biased by the excess emission in the inner disk \citep{ros12}, which is not accounted for with our simple radial power law prescription for temperature. In fitting the visibilities we cannot directly ignore the inner disk, but we can minimize its influence on the fitting process. This is accomplished in part by ignoring the channels with velocities more than 0.6 km s$^{-1}$ from the line center when comparing with the model. To minimize the central excess in the remaining channels, we also subtract off a point source, the simplest model for this emission, with a flux of 0.189$\pm$0.002 Jy, as derived from a fit to baselines $>$200k$\lambda$ across all spectral channels.  
 
 \subsection{Tests with Synthetic Data}
While, as discussed below, we find success fitting the ALMA CO(2-1) data assuming a simplified parametric model for the temperature and density structure, these parametric forms are only approximations of more detailed models \citep[e.g.][]{dal06}. In fact, high resolution observations of the disk around TW Hya \citep{deb13,and16,sch16,tea17,van17,hua18} have made it clear that, at the very least, a simple radial power law is likely not the correct model for the surface density profile. Numerous thermo-chemical models of TW Hya \citep{cal02,thi10,gor11,kam13,men14,cle15,du15} also find temperature and chemical structures that are not necessarily well matched to our simplified functional forms.

To test the ability of our modeling effort to recover the true disk structure, we fit synthetic data using our modeling prescription to see if we can recover the input model parameters. Synthetic CO data are generated based on the temperature, density, and abundance distribution from \citet{kam16}, and fed through LIME \citep{bri10} to generate model images of CO(2-1). While \citet{kam16} assume a self-similar surface density distribution, the gas temperature and CO abundance are computed without any requirement  that their radial and vertical profiles follow simple functional forms. The spectral resolution of the mock observations derived from this input model is chosen to match the ALMA observations of TW Hya's disk. Visibilities are derived from these model images at the same baselines as for the actual data, and noise is added to the visibilities based on the noise derived from the data. The synthetic, noisy visibilities are then fit as described above, with the same model prescription, the same free parameters ($q$, $R_c$, $T_{\rm atm0}$, $T_{\rm mid0}$, $\delta v_{\rm turb}$, and inclination), and the same initial parameter distribution as when fitting the actual data. We consider two sets of mock data that were generated with different turbulent velocities but identical temperature and density structure; one has weak turbulence ($\delta v_{\rm turb}$=0.05c$_s$) while the second has strong turbulence ($\delta v_{\rm turb}$=0.2c$_s$). In fitting either mock data set, we recover the input density distribution, and the gas temperature at the $\tau$=1 surface of the CO(2-1) in our best fit temperature structure is within 5\%\ of the input temperature structure. Farther from the $\tau=1$ surface, but still within the CO molecular zone, we recover the temperature only to within 20-50\%; this large uncertainty serves as a reminder that care must be taken in extrapolating the derived temperature/turbulence structure well beyond the regions contributing substantially to the emission. For the low turbulence model we measure an upper limit on the non-thermal motion of $\delta v_{\rm turb}<$0.15c$_s$, while for the high turbulence model we measure $\delta v_{\rm turb}$=0.20$\pm$0.01c$_s$, exactly as prescribed in the input model. At $R$=100 au, the high turbulence input model has $\delta v_{\rm turb}$=60 m s$^{-1}$ at the $\tau=1$ surface, while we recover $\delta v_{\rm turb}$=60$\pm$5 m s$^{-1}$, accounting for the uncertainty in both temperature and turbulence. This confirms that our modeling methodology is able to accurately recover the gas temperature at the $\tau=1$ surface, and is sensitive to strong levels of non-thermal motion, while assuming functional forms for the temperature, density, and turbulent structure that are simpler than the underlying true physical structure. 


\section{Results\label{initial_results}}
\subsection{Fiducial Model Fit}
\begin{figure*}
\center
\includegraphics[scale=.35]{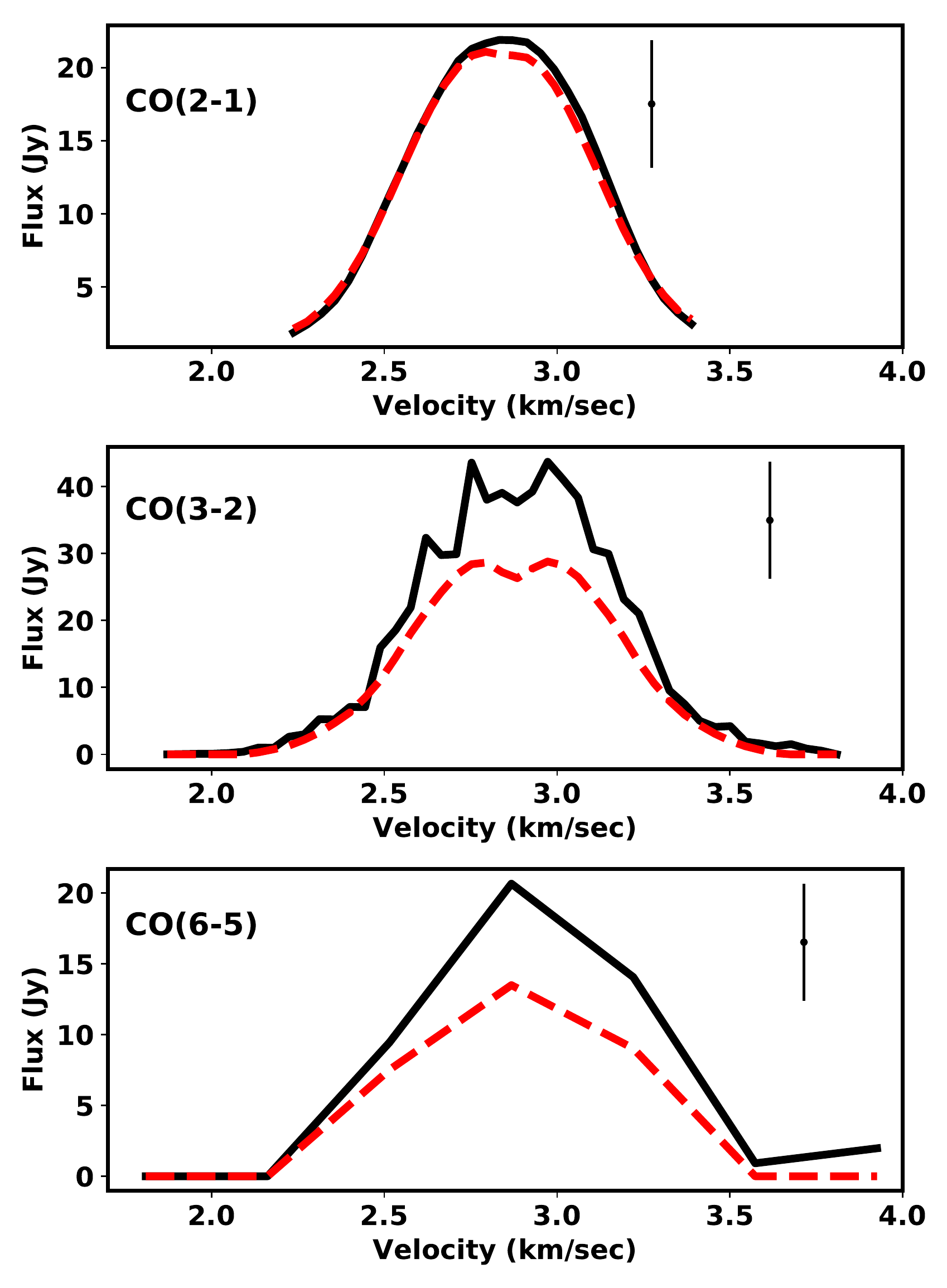}
\includegraphics[scale=.35]{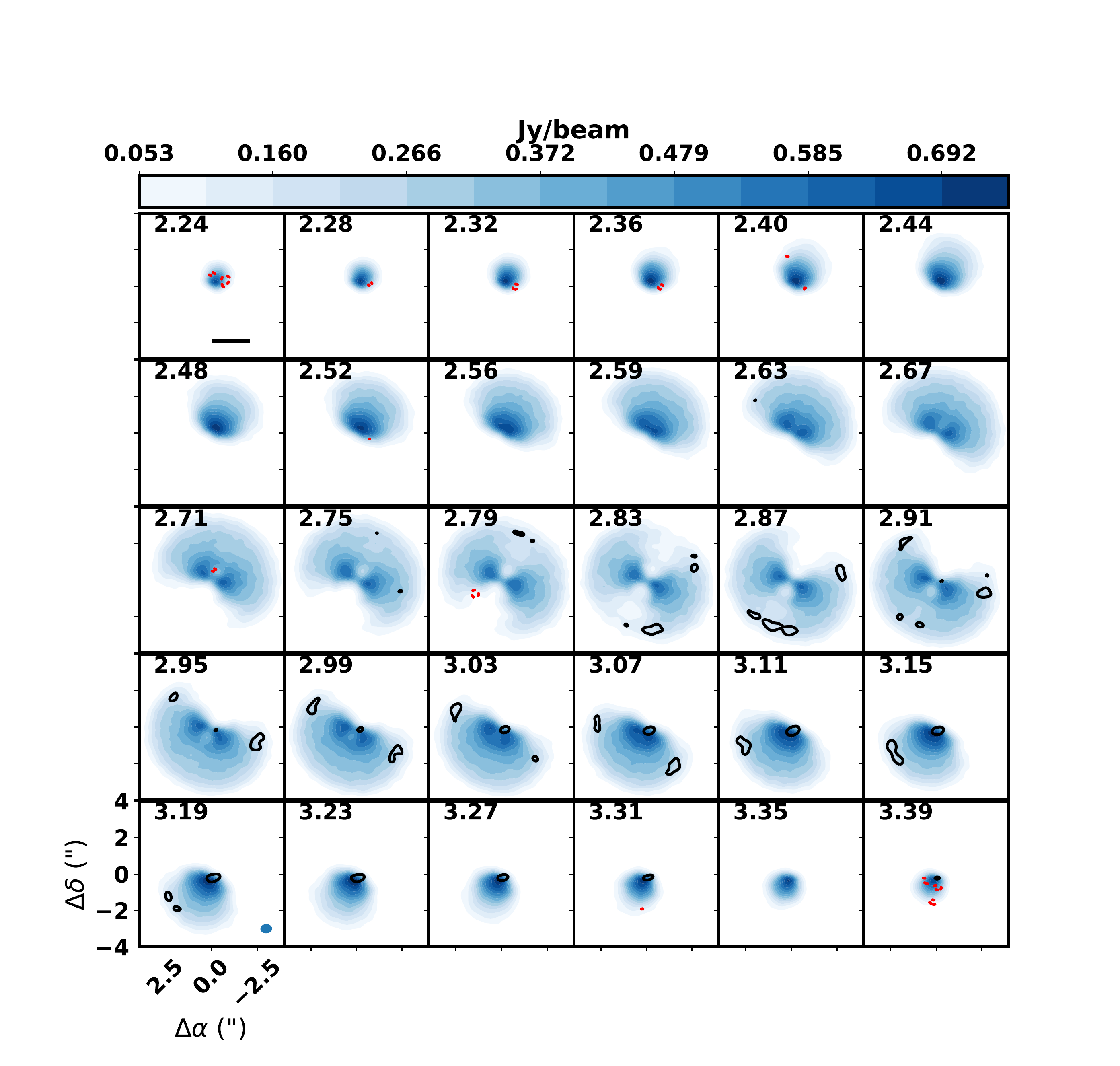}
\caption{A comparison between the fiducial model and the observed ALMA CO(2-1), SMA CO(3-2), and SMA CO(6-5) emission. The panels on the left show the spectrum, with the model marked with the red-dashed line, and the data indicated by the solid black line. A 20\%\ systematic uncertainty in the peak flux is shown by the error bar. The grid of panels on the right show the channel maps for CO(2-1), with the blue-scale showing the observed emission while the contours show the residuals, starting at 5$\sigma$ ($\sigma$=0.01 Jy beam$^{-1}$). The model spectrum closely matches the CO(2-1) data, with imaged residuals in the channel maps only reaching $\sim$7\% of the peak flux.
\label{fiducial_model}}
\end{figure*}

To analyze the ALMA CO(2-1) observations, we start with a fiducial model, allowing for the full suite of parameters to vary to find a best fit. The resulting model provides an excellent fit to the data (Figure~\ref{fiducial_model}). The model spectrum closely matches that of the data, and the residuals (generated by imaging the differenced visibilities) show only small 5$\sigma$ features at the outer edge of the disk. While these imaged residuals are statistically significant they only reach $\sim$7\%\ of the peak flux. These residuals are at a larger disk radius than the CO outer ring observed by \citet{sch16} and \citet{zha17}, but are at a similar radius as a bump in CS \citep{tea17}.  The underestimate of the disk emission by the model may reflect an underestimate of the midplane temperature, possibly due to the ability of interstellar UV radiation to penetrate the outer, lower density, regions of the disk \citep{tea17}. The fiducial model does under-predict the CO(3-2) and CO(6-5) emission at a level that is comparable to the systematic uncertainty. Additional surface heating, not included in our model but suggested by \citet{qi06} as an explanation for the elevated CO(6-5) emission, may account for some of this discrepancy. 

\begin{figure}
\center
\includegraphics[scale=.43]{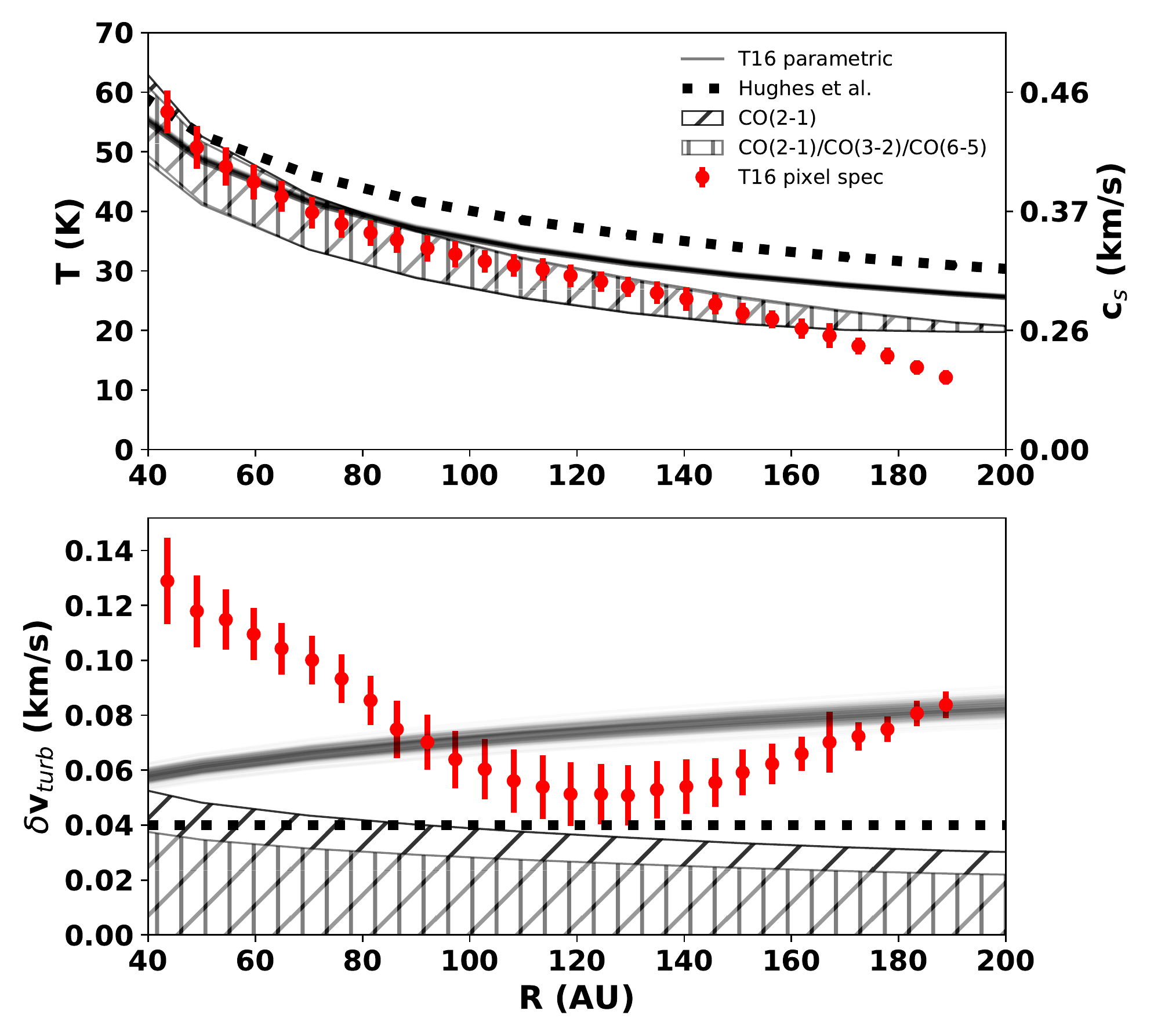}
\caption{Radial profiles of temperature (top panel), and turbulence (bottom panel). The bands mark the region in our models where 50\%\ of the CO(2-1) emission arises, as derived in our fiducial model (diagonally-lined band) and multi-line model (vertically-lined band). The single line and multi-line fit nearly completely overlap each other in terms of the temperature of the CO(2-1) emitting region. The dotted line marks the parameters derived by \citet{hug11}. The solid lines mark the structure derived by the parametric fit of T16 assuming orbital motion, temperature and turbulent line width all follow radial power laws (with its corresponding uncertainty illustrated by the thickness of the line), while the red points indicate their non-parametric fit. Throughout the disk we find a similar temperature to that of T16, with systematically lower levels of turbulence. \label{radial_compare_fid}}
\end{figure}

The derived parameters, along with their uncertainties, are listed in Table~\ref{results}. We find that we recover gas temperatures that are similar to those derived by T16, and by previous efforts to model the disk around TW Hya \citep{thi10,gor11,kam13,ber13,cle15,du15}. As discussed below, the modeled parameters (e.g. $q$, $T_{\rm atm0}$) can move slightly when assumed parameters are changed (e.g. CO abundance, $\gamma$), although these differences are small and do not strongly affect the similarity between our derived temperature structure and those of previous studies. Despite the similarity in temperature structure between our work and T16, we derive a 3$\sigma$ upper limit on the non-thermal motion ($<$0.08 c$_s$) that is lower than the 0.2-0.4 c$_s$ found by T16 (Figure~\ref{radial_compare_fid}). 

T16 point out the importance of systematic uncertainty in the absolute amplitude calibration on measurements of turbulence, and while the statistical uncertainties on the atmosphere temperature are $\sim$1\%, this does not account for systematic uncertainty in the amplitude calibration. Since CO(2-1) is optically thick over much of the disk, its flux is directly proportional to the temperature of the gas, and any uncertainty in the flux will directly translate into uncertainty on the temperature. The high S/N of these data make for very small statistical uncertainties on the flux, but large systemic uncertainties in the flux calibration still exist. T16 note a 7\%\ uncertainty in the flux of the amplitude and phase calibrator, J1037-2934. This amplitude calibrator is referenced to the absolute flux calibrator Ganymede, which itself has a $\sim$6\% uncertainty on its brightness temperature near the 230 GHz CO(2-1) line \citep{but12}. Observations of the sub-mm dust continuum emission from around TW Hya \citep{wei89,hug09,and12,qi13,hog16,nom16,and16} show variations of $\sim$10\%, consistent with uncertainties in amplitude calibration. T16 find that an absolute calibration uncertainty of 7-10\%\ limits their method to detections of turbulence $>$0.2 c$_s$ at the 5$\sigma$ level, indicating that calibration uncertainty can be the dominant source of uncertainty. Bandpass calibration uncertainties can also impart a systematic uncertainty on the derived parameters, although for ALMA the bandpass calibration is expected to be good to better than 0.2\%, and are minimized in the case of strong line emission on top of much weaker continuum emission, as is the case here. The use of identical data sets eliminates the possibility of differing calibration errors (e.g. one data set being systematically brighter than the other) as a source of concern, although the exact way in which systematic errors propagate through our method and those of T16 may still contribute to the discrepancy in our results. We first directly consider the role of systematic amplitude calibration uncertainty on our model fits and then we fit CO(2-1), CO(3-2), and CO(6-5) simultaneously, utilizing the ability of the flux ratios of these three lines to help mitigate the influence of calibration uncertainties on our measurement of the gas temperature. 

\subsection{Amplitude Calibration Uncertainty}
\begin{figure}
\includegraphics[scale=.4,angle=270,origin=c]{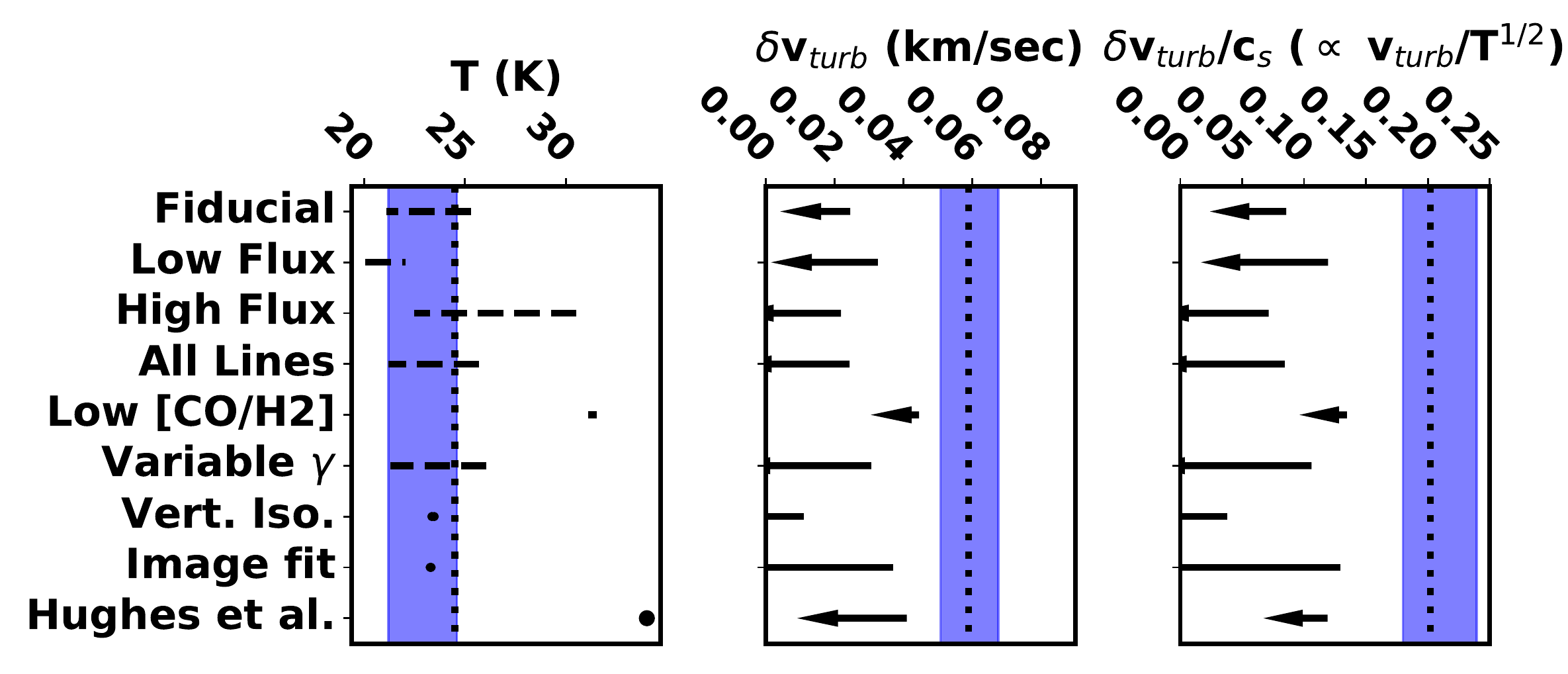}
\caption{A comparison between the derived temperature and turbulence (both in units of km/sec, and in units of the local sound speed) at R=150 au for different models. The dashed lines in the temperature panel are used for models with a vertical temperature gradient, and they mark the range of temperatures that contribute to the emission at this radius. For all other points the extent of the line indicates the 3$\sigma$ range of the posterior distributions. The blue band indicates the $\pm$1$\sigma$ range around the T16 results, while the dotted line is an independent implementation of T16's method as applied to our CO(2-1) channel maps. Across all of the model prescriptions that we have explored we consistently measure weak turbulence, suggesting that this result is not strongly model dependent. \label{model_compare}}
\end{figure}

To determine the robustness of our result against systematic uncertainty in the amplitude calibration, we rerun our fitting with the model visibilities either scaled up or down by 20\%\ before comparing them to the data. By scaling the model visibilities upward/downward by 20\% the model is forced to become fainter/brighter by 20\% in order to fit the data. The uncertainty on the amplitude calibrators are 3-15\%\ \citep{but12}, but we choose a systematic uncertainty value of 20\% as an upper limit on the potential size of the actual uncertainty. The results of both the low and high flux fit to the data are listed in Table~\ref{results}; we find that systematic uncertainty strongly affects the derived temperature, but not the resulting turbulence, which is still parameterized as proportional to the local sound speed. The high and low flux models cause the temperature in the CO(2-1) emitting region to vary by $\sim$30\% (Figure~\ref{model_compare}) while the turbulence limits vary from $<$0.12c$_s$ ($<$0.03 km s$^{-1}$ at $R$ = 150 au) in the low flux case to $<$0.07c$_s$ ($<$0.02 km s$^{-1}$ at $R$ = 150 au) in the high flux case. This similarity in turbulence arises because we utilize the full spatial information of the 3D position-position-velocity cube to help constrain the turbulence; as discussed in \citet{sim15}, turbulence increases the spatial broadening of the emission in individual channel maps making the surface area of the emission dependent on turbulence. Temperature also influences the spatial broadening of the emission, but the ranges of temperatures allowed by the data, and hence the amount of spatial broadening that can be explained by thermal motion, is limited by the noise, in particular the amplitude calibration uncertainty, on the surface brightness of the emission. Turbulence does not have such a strong ancillary constraint, and has more freedom to move in response to the surface area of the emission. The surface area of the emission is not influenced by the scaling of the flux, making our turbulence limit less sensitive to the amplitude calibration. 

One key aspect of using the spatial broadening is that our models self-consistently, through the hydrostatic equilibrium calculation, vary the density structure in concert with changes to the temperature. As the gas temperature increases, in particular the midplane temperature, the disk becomes puffier, pushing the $\tau$=1 surface to higher altitudes in the disk, where the gas is warmer. This amplifies the change in brightness associated with a given change in gas temperature, resulting in e.g. a 12\%\ increase in brightness for a 10\%\ increase in gas temperature among the models with a vertical temperature gradient. T16 model the CO density structure as independent of the temperature structure, potentially leading to a stronger degeneracy between thermal and non-thermal broadening. 


\subsection{Multi-line fitting}
One way to circumvent the temperature uncertainty associated with amplitude calibration is to fit multiple transitions of a single molecule. For emission in local thermodynamic equilibrium, the relative level populations of different transitions of a single molecule are set by a single excitation temperature. But the direct use of relative emission strengths to measure the excitation temperature assumes that the emission from all three lines arises from gas of the same temperature. Differences in the optical depth between the three lines can lead them to probe gas at different heights, and hence different temperatures, within the disk. Within our modeling framework this complication is automatically accommodated through the use of a single temperature and density structure to self-consistently derive the flux of the different emission lines, allowing for different excitation temperatures for transitions that arise from different heights within the disk.

We perform a multi-line fitting, including CO(2-1), CO(3-2), and CO(6-5) using the same model prescription as in the fiducial model. We find a nearly identical result between the multi-line fit and the single line fit (Table~\ref{results}, Figure~\ref{radial_compare_fid}). The result is a marginally significant improvement in the fit to CO(3-2), with no significant difference in the fit to the noisy CO(6-5) observations. While this fit is driven in large part by the low noise and substantial uv-plane sampling of the ALMA data, our ability to self-consistently fit multiple CO transitions with a single disk structure confirms that the temperature profile is a reasonably accurate representation of the disk around TW Hya. As with the fiducial model we constrain the turbulence to $<$0.08c$_s$. 




\subsection{CO abundance}
\citet{yu17} recently suggested that using the peak-to-trough ratio of the CO emission line spectral profile as a measure of turbulence, without accounting for a decrease in CO abundance below the canonical [CO/H$_2$]=10$^{-4}$ value, or the steepening in the CO abundance radial profile with time as carbon becomes sequestered into complex species \citep{yu16}, would lead to an underestimate of the non-thermal motion. We can address this potential degeneracy in the context of our modeling framework. We first consider a model with CO abundance dropped to 10$^{-7}$, instead of the fiducial abundance of 10$^{-6}$, to mimic an evolution in the CO abundance. An abundance of 10$^{-7}$ is close to a lower limit on the actual abundance of CO within TW Hya; below this level the CO(2-1) emission becomes optically thin throughout much of the disk and it becomes impossible to reproduce the large observed flux without e.g. increasing the total gas mass to a value that is nearly equal to the stellar mass. 

We find that our limit on turbulence increases slightly, from $<$0.08c$_s$ in the fiducial model to $<$0.13c$_s$ in the low abundance model (Table~\ref{results}, Fig~\ref{model_compare}). This corresponds to a difference in non-thermal motion of $\sim$20 m s$^{-1}$ at 150 au, smaller than the $\sim$100 m s$^{-1}$ error predicted by \citet{yu17}. We note that the low abundance model requires a midplane temperature that increases with radius ($T_{\rm mid}\sim r^{0.25}$), a nearly vertically isothermal structure ($T_{\rm atm0}=32.2^{+0.7}_{-0.6}$ K, $T_{\rm mid0}=31.2^{+0.4}_{-0.3}$ K) in the outer disk, under-predicts the optical depths of $^{13}$CO(3-2) and $^{13}$CO(6-5) as measured by \citet{sch16}, and results in a significantly worse fit to the data. 

We additionally test for a steepening of the radial gradient of CO abundance by allowing $\gamma$, the power law index on the surface density profile, to vary. This approximates a change in the CO abundance profile given that a change in the shape of the surface density profile will lead directly to a change in the shape of the CO abundance profile. Varying the surface density profile will introduce additional effects (e.g. changing the perturbation to the orbital motion caused by the pressure gradient) that are not present when simply varying the CO abundance, but over the range of parameters considered here these differences are expected to be negligible. We again find a turbulence limit ($<$0.09c$_s$) that is consistent with our fiducial model ($<$0.08c$_s$), which is not surprising since the derived range on $\gamma$ (=0.94$^{+0.09}_{-0.12}$) encompasses our fiducial value of $\gamma$=1 (Table~\ref{results}, Fig~\ref{model_compare}). We also consider a model in which $\gamma$ is fixed at a steep value ($\gamma$=1.9) but are unable to find an acceptable fit to the data under this condition. The model defined by the peak of the posterior distributions underestimates the peak spectral line flux by $\sim$20\%, with 20-30\%\ residuals in the individual channel maps. While the limit on turbulence has risen to $<$0.22c$_s$, the poor quality of the fit suggest that we cannot interpret this as a realistic limit on the turbulence.  

The minimal influence of a mis-estimated CO abundance within our model fitting is due to the fact that we utilize the full three-dimensional data set, instead of simply the peak-to-trough ratio of the spectrum. By fitting the visibilities across all spectral channels we are leveraging all of the available information to minimize any degeneracies with e.g. CO abundance. Previous efforts to measure turbulence \citep{car04,hug11,gui12,fla15,tea16,fla17}, utilize, at minimum, the full spectral line profile in searching for non-thermal motion and can rely on differences in e.g. the line wings \citep{yu17} to limit the degeneracy between CO abundance and turbulence. Additionally, in \citet{fla17}, when fitting DCO$^+$(3-2) and C$^{18}$O(2-1), we explicitly include free parameters for the DCO$^+$ and CO abundance respectively and find that it does not strongly bias our ability to constrain turbulence to $<$0.05c$_s$. \citet{fla17} also point out, in the context of fitting the CO(2-1) emission from around HD 163296, that the full three-dimensional data set may prefer a low value of turbulence even when this produces a model that does not match the observed peak-to-trough ratio. Our tests in the context of TW Hya also bear out the result that assumptions about the CO abundance profile do not strongly bias our turbulence limits.

\subsection{Vertically Isothermal vs Vertically Non-isothermal}
Protoplanetary disks, being dominated by heating from the central star, are expected to have a strong vertical gradient in temperature, with the warmest gas at the surface layers and the coldest gas near the midplane \citep[e.g.][]{dal06}. We have chosen a particular functional form to represent this gradient, which itself is based on more detailed radiative transfer models. While observational evidence in the sub-mm/mm confirms the presence of a temperature gradient in the outer disk \citep[e.g.][]{pie07,pan08,pan10,ros13,dut17}, analysis of prior observations have not yet constrained the exact shape of this gradient.

To understand the role of the vertical temperature structure in our model fitting, we consider a vertically isothermal model. T16, in both their pixel-by-pixel and parametric approach, assume that the disk is vertically isothermal. We also parameterize turbulence directly in units of km s$^{-1}$ with a power law radial profile ($\delta v_{\rm turb}$=$\delta v_{\rm turb0}$(r/150 au)$^{q_{\rm turb}}$) to bring our model in closer agreement with the structure in T16. The results from this model fit are listed in Table~\ref{results}. We find a significantly lower $T_{\rm atm0}$ as compared to the fiducial model, which is due to a difference in how these parameters relate to the temperature of the $\tau$=1 surface. In the presence of a vertical temperature gradient, the excitation temperature of CO(2-1) is always less than or equal to $T_{\rm atm}$ at a given radius, while within a vertically isothermal disk the excitation temperature of CO(2-1) is equivalent to $T_{\rm atm}$. Figure~\ref{model_compare} demonstrates that despite the difference in the $T_{\rm atm0}$ parameter between the isothermal and non-isothermal models they both result in the same gas temperature within the CO(2-1) emitting layer (both models have CO emission arising from 2-3 pressure scale heights above the midplane).

The constraint on turbulence has also varied between the non-isothermal and isothermal models, with the vertically isothermal model limiting the non-thermal motion to less than 10 m s$^{-1}$ ($\lesssim$0.04c$_s$ in the outer disk), below the limit found from the vertically non-isothermal model ($<$0.08c$_s$, corresponding to $\lesssim$24 m s$^{-1}$ at $R$=150 au). This significant drop in turbulence may be due to the warmer midplane emission in the vertically isothermal model (23.42$^{+0.03}_{-0.08}$ K) as compared to the vertically non-isothermal model (15$\pm$2 K). Along the edges of the emission in individual channels, the lines of sight have low optical depth and reach the midplane (Figure~\ref{tau1_figure}). In the non-isothermal model, increasing turbulence will allow more molecular emission to 'bleed' over from nearby channels, increasing the optical depth along these edge lines of sight, leading to emission from higher, and warmer, regions of the disk. The result is higher flux along the edges of the emission for larger turbulence levels. With the warm midplane in the vertically isothermal model, the emission along these lines of sight will be high enough to match the observations without the need for high turbulence. As a result, with the warm midplane in the vertically isothermal structure an acceptable fit can be found with very weak turbulence. In fitting the mock $\delta v_{\rm turb}$=0.2c$_s$ model observations with a vertically isothermal disk we find that we derive low levels of turbulence ($\delta v_{\rm turb}\lesssim$0.1c$_s$), similar to the effect seen when fitting the TW Hya data. 

Recent observations of $^{13}$C$^{18}$O(3-2) and C$^{18}$O(3-2) indicate that the CO around TW Hya may freeze-out at 27 K instead of the 19 K assumed in our fiducial model \citep{zha17}, suggesting that the midplane should be warmer than assumed in our fiducial vertically non-isothermal model. Given that the vertically isothermal models indicate that the inclusion of a warm midplane drives down our estimate of turbulence, we anticipate that the freeze-out temperature assumed in our fiducial model results in a more conservative upper limit on the turbulence. 


\subsection{Image Plane Fitting}
Turbulent broadening has also been derived by fitting emission line spectra in the image plane. We quantify the effect of this on turbulence by using the same vertically isothermal model employed in the previous section and instead of converting model images to model visibilities, we convolve the model images with a Gaussian with the same FWHM as the observed beam and compare directly with the channel maps. The noise is estimated based on the RMS within a region well separated from the disk emission. We exclude pixels within 40 au of the disk center in the image plane, matching the procedure in T16. We also use the image presented in T16, binned to the same spectral resolution as in the rest of our analysis, to eliminate any differences in imaging and clean deconvolution parameters that may confound our result. 

We find that fitting to the image plane produces nearly identical results to fitting the visibilities. The disk temperatures are nearly identical ($T_{\rm atm0}$=23.42$^{+0.03}_{-0.08}$ in the visibility fit vs 23.3$^{+0.2}_{-0.1}$ in the image plane fit) while the turbulence limits are consistently smaller than found in T16. The consistency between the image plane fit and the visibility domain fit suggests that this is not the source of the discrepancy between our result and T16.

In general, since the primary data products from a radio interferometric telescope are the visibilities and not the images, caution must be used when fitting to the image plane. In particular, the uncertainties in the image plane may not be an accurate measure of the uncertainties on the data. Typically the uncertainty in an image is estimated based on the RMS in an emission free region. This commonly used method suffers from the fact that the cleaning process is not a linear transformation of the visibilities, and hence the uncertainties on the (directly-measured) visibilities are not accurately transformed into the image plane. Additional artifacts due to the cleaning process, which will vary depending on the exact choice of cleaning depth, cleaning area, robust parameter, etc., can also contribute to the RMS, as well as change the exact emission that is being modeled. This can be especially problematic for high S/N data for which cleaning artifacts are comparable to, or even larger than, the noise in the data. 

The image RMS also does not account for systematic uncertainties on the recovered flux. Interferometric observations must have sufficient short-baseline coverage to ensure that the total flux is recovered. The minimum uv distance for the ALMA observations are $\sim$15k$\lambda$, which corresponds to a largest resolvable angular scale of $\sim$14$\arcsec$. This is well beyond the edge of the disk emission, which only extends over a diameter of 8''. \citet{ros12} measure an integrated flux of 17.5$\pm$1.8 Jy km s$^{-1}$, with a similar minimum baseline as our observations, which is consistent with our 16.3$\pm$3.3 Jy km s$^{-1}$. 

The ability to recover the distribution of flux on the sky can be quantified using the image fidelity, defined as the ratio of the true flux to the difference between this true flux and the image reconstructed from the visibilities. In practice the true flux of an observed object is unknown, making image fidelity impossible to estimate. We can perform this calculation here using the mock high turbulence observations used to test our ability to recover the input temperature and turbulence distribution. In the case of the TW Hya observations, and assuming no noise on the visibilities, we can recover, on average, 98\%\ of the input flux, although this can range from 68\%\ to 99.97\%\ across the entire data set. In the case of the high S/N observations presented here, including noise on the visibilities does not substantially change this distribution, indicating the image reconstruction is limited more by the antenna configuration (e.g. missing short baselines, or poorly sampled long baselines) than noise in the data.

Even with high image fidelity, the RMS may under-estimate the noise on the data. Pixels within a single beam are not independent of one another, and errors derived from the RMS do not account for the covariance between these nearby pixels. We can estimate the strength of the covariance by randomly generating a series of images from the same visibility data set, where the real and imaginary part of each visibility point is drawn from a Gaussian distribution defined by the uncertainty on that visibility point, and comparing the relative variation in flux among nearby pixels in the resulting cleaned images. We find that among adjacent pixels the covariance is equal in strength to the variance of a given pixel, and even pixels 1/2 a beam apart have a covariance that is half as large as the variance. Any model fitting must take into account the additional source of uncertainty associated with the covariance in neighboring pixels.

The inter-dependence of nearby pixels can also lead to over-fitting of the data if each pixel within a single beam is fit independently. For example in the TW Hya ALMA CO(2-1) channel maps each spatial pixel has $\sim$10-20 spectral channels of independent information, which can be fit with three parameters (optical depth, temperature, turbulence). But, there are $\sim$20 pixels within the FWHM of a single beam, and if each pixel is fit independently then $\sim$60 parameters have been used to fit the $\sim$10-20 independent spectral channels within a single beam. Any pixel-to-pixel difference that is due to noise will be fit as real variations in the optical depth/temperature/turbulence across the beam, leading to an under-estimate in the uncertainty on the physical parameters. 

Additional biases can occur when comparing an un-convolved model with a convolved data set. In the case of TW Hya, a single beam corresponds to $\sim$30 au across the disk and the temperature/density/turbulence will all vary across this width. We have tested this effect by repeating our image plane fitting while excluding the convolution step. We find nearly identical results when excluding convolution suggesting that, as pointed out in T16, the beam size is small enough in these data that convolution does not strongly influence the results. Given the shape of the temperature radial profile, the gas temperature is only expected to vary by $\sim$5\%\ across a single beam. While convolution does not play a substantial role here, it can play a larger role in lower resolution observations of other systems, or in cases where certain features (e.g. rings) have sizes comparable to or smaller than the spatial resolution of the data. 

In general, while we find similar solutions when fitting either the image plane or the visibilities in the case of TW Hya, caution must be used in interpreting results derived entirely from image plane fitting. Image fidelity will depend strongly on the sampling of the uv-plane, the comparison of a convolved data set with an un-convolved model will become more problematic for poorer linear spatial resolutions, and the covariance of neighboring pixels will grow as the noise increases. The similarity between fitting the image plane and fitting the visibilities that is found here may be unique to this system (TW Hya) and to the parameter (turbulence) under consideration.



\section{Discussion}
\subsection{Comparison of Methods}
We have examined in detail the role of amplitude calibration uncertainty, assumptions about the CO abundance, the assumed functional form of the vertical temperature structure, and fitting in the image plane in potentially biasing our ability to tightly constrain turbulence in the disk around TW Hya. We find that the vertical temperature gradient can influence the derived turbulence, and fitting in the image plane introduces a host of complicating factors. The role of the temperature profile in constraining turbulence is not surprising given that temperature and turbulence both contribute to the line broadening; identical line broadening can be achieved by increasing the gas temperature while decreasing the turbulence, or vice versa. This degeneracy is especially problematic if the functional form of e.g. temperature provides it more freedom to vary across the disk, potentially leading the models to prefer temperature in place of turbulence simply due to its greater flexibility. If, for example, turbulence within the disk varied radially, but the parameterization of turbulence assumed a constant value throughout the disk, then the models may use the radial variation in the temperature structure to fit the line broadening in place of significant turbulence. Effects on the emission that extend beyond line broadening can help to disentangle thermal and non-thermal motion. T16, in their pixel-by-pixel approach, utilize the relationship between flux and temperature for an optically thick line to derive the thermal broadening, which necessitates the assumption of a single temperature along a single line of sight through the disk. We assume a coupling between the vertical temperature and density structures due to hydrostatic equilibrium, while T16, in their parametric models, treat temperature and density as independent factors. As discussed in the context of the amplitude calibration uncertainty, this coupling can amplify the effects of a given change in gas temperature on the emission profile (e.g. a 10\% change in gas temperature results in a $>$10\% change in flux), making it easier to distinguish between the effects of temperature and turbulence. 

Our models also differ from T16 in how the CO is distributed vertically within the disk. In their parametric model, T16 assume that the CO density distribution is well approximated by a Gaussian shape in the vertical direction, centered at the midplane, while we assume a constant CO abundance within a molecular layer whose upper and lower boundaries are defined by photodissociation and freeze-out. The constant CO abundance, when combined with the vertically and radially varying gas density, produce the CO gas density distribution used in the radiative transfer calculation. Recent ALMA observations have spatially resolved the CO molecular layers within the disks around HD 163296 \citep{ros13,deg13}, HD 34282 \citep{vanp17}, IM Lup \citep{pin17}, and the edge-on 'Flying Saucer' \citep{dut17} supporting the use of a layered prescription for the CO gas. The exact details of this prescription (e.g. freeze-out below 19 K) may differ from those prescribed within our model; \cite{zha17} find a CO freeze-out temperature of 27 K around TW Hya and \citet{qi15} find evidence for a CO freeze-out temperature of 25 K around HD 163296, as well as a CO depletion in the freeze out zone that is not as severe as we assume here. 

Our assumptions about the coupling between temperature and density (through hydrostatic equilibrium) and the CO abundance distribution are physically motivated, and our ability to accurately recover the temperature and turbulence at the CO(2-1) $\tau=1$ surface from mock observations lends credence to our methodology. But the results of fitting these mock observations begin to deviate strongly from the underlying structure when extrapolating well beyond the CO emitting regions within the disk, and we employed mock observations whose underlying turbulence structure, as a constant fraction of the sound speed throughout the disk, exactly matches our input prescription. Further refinements to the underlying model prescription, and the associated model assumptions, can be made by using observations that provide model-independent measures of the temperature and density in the outer disk. Spatially resolving the molecular emission layers above and below the midplane \citep{ros13,deg13,vanp17,pin17,dut17} provides a geometric constraint on the midplane temperature, which influences the puffiness of the disk and the distance between the molecular layers. Observations of molecules with hyperfine emission structure such as methyl cyanide \citep{obe15}, cyanide \citep{tea16}, and ammonia \citep{sal16}, can be used to directly measure the gas temperature, independent of amplitude calibration uncertainty. A mix of optically thin and optically thick tracers can directly constrain the temperature and density of the disk \citep[e.g.][]{sch16,zha17}. Observations of dust settling \citep{mul12,bon16,pin16,dej16}, and mechanical heating \citep{naj17}, while not directly measuring the motion of the gas, can still constrain the turbulence through its secondary effects. Combinations of observations can then provide the best prospect for constraining such small effects as turbulence.



\subsection{Implications of modest turbulence around TW Hya\label{theory}}
We have found that turbulence in the outer disk around TW Hya is limited to $<$0.08c$_s$ ($\lesssim$30 m s$^{-1}$ at 150 au) localized to the CO emitting layer of the disk. Similarly weak non-thermal motions were found in our molecular line observations of gas around HD 163296 \citep[$v_{\rm turb}<0.04$c$_s$][]{fla15,fla17}. A primary quantity of interest in all theoretical studies of disk turbulence is the so-called alpha parameter, which is defined relative to the viscosity, $\nu$, as $\nu=\alpha c_s H$ \citep{sha73}, where $H$ is the pressure scale height of the disk. In numerical simulations $\alpha$ is calculated as a mass-weighted quantity, making it strongly weighted toward the high density midplane where, in the context of the MRI, the turbulence is predicted to be weaker \citep[e.g.][]{sim15}. Our TW Hya limit corresponds to $\alpha<0.007$ but applies directly to the CO(2-1) emitting region, which according to our models lies between two and three pressure scale heights above the midplane.
Rather than examine global values of $\alpha$, we can more accurately compare the expected size of non-thermal motions in the CO emitting layers in the disk atmosphere. Our upper limit for TW Hya is inconsistent with the strongest predictions of turbulence, but is consistent with the modest levels of turbulence found in the upper disk layers for models with weak UV ionization and/or weak magnetic fields \citep{bai15,sim15,sim17}. Weak UV ionization will limit the regions of the disk that can become sufficiently coupled to the magnetic field, and hence are susceptible to the MRI. \citet{per11} find that FUV ionization can penetrate down to vertical column densities ranging from $\Sigma$=0.01 g cm$^{-2}$ to 0.1 g cm$^{-2}$, depending on the grain population and FUV luminosity, while the CO(2-1) emission around TW Hya originates from near 0.1 g cm$^{-2}$. If the ionization field were able to penetrate down to $\Sigma=0.1$ g cm$^{-2}$ we would anticipate measuring strong turbulence, but ionization that only reaches $\Sigma=0.01$ g cm$^{-2}$ would result in a weaker turbulence, consistent with our observations. \citet{fra14} find that the FUV luminosity originating from near TW Hya's stellar surface is not significantly weaker than other T Tauri stars, but in order for this emission to ionize the outer disk it must first reach the outer disk. A wind, which is suggested by gas observations \citep{pas09,pas11,pas12}, could shield the outer disk from FUV photons that originate from near the star. \citet{cle15} use ALMA observations of molecular emission to directly measure the ionization of the gas near the midplane of the outer disk, and find that the observations are consistent with cosmic ray ionization that is orders of magnitude weaker than traditionally assumed. Weak ionization in the upper disk layers of the outer disk may be consistent with the lack of turbulence at the CO(2-1) emitting layers in the disk. However, numerical simulations that probe the physics of such weak ionization in the presence of magnetic fields will be needed to test this consistency; such work is underway \citep{sim17}. The weak turbulence may also be consistent with purely hydrodynamic instabilities, such as the Vertical Shear Instability \citep{flo17}.



Weak turbulence is also consistent with the low accretion rate onto the star, and the advanced age of the disk. The accretion rate through the disk is related to the surface density and $\alpha$ through:
\begin{equation}
\alpha = \frac{\dot{M}\Omega}{3\pi\Sigma c_s^2}(1-\sqrt{r_{in}/r}),
\end{equation}
where $\dot{M}$ is the accretion rate, $\Omega$ is the Keplerian orbital frequency, $\Sigma$ is the surface density, $c_s$ is the local sound speed, and $r_{in}$ is the inner radius of the disk \citep{arm11}. Recent observations of HD emission \citep{ber13}, as well as CO isotopologues \citep{sch16,zha17}, provide accurate estimates of the gas surface density and temperature structure in the outer disk. When combined with our upper limit on $\alpha$ we derive accretion rates through the outer disk of $<$4$\times$10$^{-9}$M$_{\odot}$ yr$^{-1}$. This assumes that our limit on $\alpha$ is representative of the turbulent behavior throughout the disk, which is a reasonable assumption given that MRI predicts that the non-thermal motion will be weaker at the midplane than in the upper disk layers probed by our observations. An accretion rate through the outer disk near the predicted upper limit would deplete the disk in 12.5 Myr, similar to the $\sim$10 Myr age of the system. This accretion limit is also consistent with the measured accretion rate onto the star \citep[2$\times$10$^{-9}$ M$_{\odot}$ yr$^{-1}$;][]{ale02,her04,ing13}. 


\section{Conclusions}
Turbulence, whether it be from the magneto-rotational instability, the vertical shear instability, etc, is a fundamental component of the growth and evolution of planets in circumstellar disks around young stars, but its predicted strength is modest, making robust detections a challenging proposition. In analyzing CO(2-1) emission from around TW Hya, across multiple different model prescriptions we find a smaller upper limit on turbulence ($\delta v_{\rm turb}<$0.08c$_s$), with a range of $<$0.04c$_s$ - $<$0.13c$_s$ depending on the exact model prescription. All of these fall below the turbulent broadening inferred by T16 ($\delta v_{\rm turb}\sim0.2-0.4$c$_s$). This discrepancy, when expressed in absolute velocity units, is small ($\sim$30 m s$^{-1}$ at 150 au), but important, as it represents the difference between a modestly turbulent $\alpha\sim$0.09 and a weakly turbulent $\alpha<$0.007. Such weak turbulence is consistent with numerical models of the magneto-rotational instability with weak UV ionization and/or weak magnetic fields \citep{bai15,sim15,sim17}.

We consider the influence on measurements of turbulence of uncertainties in the amplitude calibration, assumptions about the CO abundance, the functional form of the vertical temperature structure, and fitting in the image plane. While the vertical temperature structure and fitting in the image plane can play an important role in accurately constraining turbulence, and result in -50\% and +63\% variations on the turbulence limit respectively, the source of the discrepancy between our result and T16 likely lies in the coupling between temperature and density through hydrostatic equilibrium within our models, and/or the fact that we distribute CO within a molecular layer above the midplane. Using hydrostatic equilibrium to calculate the vertical density structure and confining CO to a molecular layer bounded by photodissociation and freeze-out are well motivated physically and observationally, but the prescriptions employed here may not apply exactly throughout the entire disk, and further work is needed to determine the exact effect of these assumptions on measurements of turbulence. Complementary observations of e.g. hyperfine emission sensitive to temperature, can provide model-independent constraints on the temperature and density structure of the disk, providing further refinements to the model prescription, and ensuring robust constraints on the weak non-thermal motion.

\acknowledgements
We thank C. Qi for providing use with the CO(6-5) data. KF and AMH are supported in part by NSF grant AST-1412647. This paper makes use of the following ALMA data: ADS/JAO.ALMA\#2013.1.00387.S. ALMA is a partnership of ESO (representing its member states), NSF (USA) and NINS (Japan), together with NRC (Canada), NSC and ASIAA (Taiwan), and KASI (Republic of Korea), in cooperation with the Republic of Chile. The ALMA Observatory is operated by ESO, AUI/NRAO and NAOJ. The National Radio Astronomy Observatory is a facility of the National Science Foundation operated under cooperative agreement by Associated Universities, Inc. The research made use of Astropy, a community-developed core Python package for Astronomy \citep{astropy13}. We thank Wesleyan University for time on its high performance computing cluster supported by the NSF under grant number CNS-0619508. The Submillimeter Array is a joint project between the Smithsonian Astrophysical Observatory and the Academia Sinica Institute of Astronomy and Astrophysics and is funded by the Smithsonian Institution and the Academia Sinica. 

\clearpage

\begin{deluxetable}{ccccccccc}
\tabletypesize{\scriptsize}
\tablewidth{0pt}
\tablecaption{TW Hya model results\label{results}}
\tablehead{ \colhead{Description} & \colhead{$q$} & \colhead{$\log (R_c (\rm au))$} & \colhead{$T_{\rm atm0}$ (K)} & \colhead{$T_{\rm mid0}$ (K)} & \colhead{inclination ($^{\circ}$)} & \colhead{$\delta v_{\rm turb}$} & \colhead{$q_{\rm turb}$} & \colhead{$\chi^2_{\nu}$}}
\startdata
Fiducial & -0.46$^{+0.02}_{-0.03}$ & 1.61$^{+0.09}_{-0.06}$ & 33$\pm1$ & 15$\pm2$ & 6.2$\pm$0.1 & $<$0.08c$_s$ & n/a & 1.0306\\ 
\hline
Low Flux & -0.52$\pm0.03$ & 1.84$^{+0.07}_{-0.06}$ & 25$\pm1$ & 14$^{+2}_{-1}$ & 6.0$\pm0.1$ & $<$0.12c$_s$ & n/a & 1.0286\\ 
High Flux & -0.35$^{+0.06}_{-0.04}$ & 1.59$^{+0.05}_{-0.03}$ & 46$^{+5}_{-3}$ & 14$^{+2}_{-3}$ & 6.2$\pm0.1$ & $<$0.07c$_s$ & n/a & 1.0338\\ 
 All lines & -0.44$\pm0.01$ & 1.61$^{+0.02}_{-0.04}$ & 33.5$^{+0.4}_{-0.6}$ & 15.0$^{+1.4}_{-0.6}$ & 6.14$^{+0.05}_{-0.03}$ & $<$0.06c$_s$ & n/a & 1.0304\\ 
 Low [CO/H$_2$] & -0.77$\pm0.01$ & 1.623$^{+0.002}_{-0.001}$ & 32.2$^{+0.7}_{-0.6}$ & 31.2$^{+0.4}_{-0.3}$ & 6.03$^{+0.03}_{-0.01}$ & $<$0.13c$_s$ & n/a & 1.0341\\ 
 Variable $\gamma$\tablenotemark{a} & -0.43$^{+0.04}_{-0.06}$ & 1.67$^{+0.05}_{-0.08}$ & 34.2$\pm$3 & 14$^{+3}_{-2}$ & 6.2$^{+0.1}_{-0.2}$ & $<$0.09c$_s$ & n/a & 1.0306\\
\hline
 Vertically isothermal & -0.471$^{+0.003}_{-0.001}$ & 1.433$\pm0.001$ & 23.42$^{+0.03}_{-0.08}$ & n/a & 6.77$^{+0.02}_{-0.01}$ & $<$10 m s$^{-1}$\tablenotemark{b} & -0.1$^{+0.2}_{-0.4}$ & 1.0573\\ 
 Image plane & -0.39$\pm0.01$ & 1.418$^{+0.001}_{-0.002}$ & 23.3$^{+0.2}_{-0.1}$ & n/a & 6.40$^{+0.01}_{-0.03}$ & $<$36 m s$^{-1}$\tablenotemark{c} & 0.03$^{+1.22}_{-0.86}$ & 1.4208 \\ 
\hline
T16 parametric results\tablenotemark{d} & -0.464$\pm$0.003 & n/a & 29.3$\pm$0.6 & n/a & 7 & 65$\pm$6 m s$^{-1}$\tablenotemark{e} & -0.22$\pm$0.03 & \ldots\\ 
 Hughes et al. & -0.4 & n/a & 40 & n/a & 7 & $<$40 m s$^{-1}$\tablenotemark{f} & n/a & \ldots\\
\enddata
\tablecomments{Median, plus 3$\sigma$ ranges, of the marginalized posterior distribution functions.}
\tablenotetext{a}{The power law exponent on the surface density profile, $\gamma$, is allowed to vary in addition to the other parameters. The resulting median, plus 3$\sigma$ range, for $\gamma$ is 0.94$^{+0.09}_{-0.12}$}
\tablenotetext{b}{$<$0.04c$_s$ at R=150 au}
\tablenotetext{c}{$<$0.13c$_s$ at R=150 au}
\tablenotetext{d}{T16's parametric model fit assuming that orbital motion, temperature, and turbulent linewidth follow radial power laws. Similar results are found if the total linewidth, instead of the non-thermal motion, is assumed to follow a radial power law, and if each pixel is fit independently.}
\tablenotetext{e}{0.20$\pm$0.02c$_s$ at R=150 au}
\tablenotetext{f}{$<$0.10c$_s$ at R=150 au}
\end{deluxetable}

\end{document}